\pdfoutput=1

\documentclass[11pt]{article}

\usepackage[final]{nlp4MusA}

\usepackage{times}
\usepackage{latexsym}

\usepackage[T1]{fontenc}

\usepackage[utf8]{inputenc}

\usepackage{microtype}

\usepackage{inconsolata}

\usepackage{graphicx}

\usepackage[disable]{todonotes} 

\usepackage{tikz}
\usepackage{xspace}
\usepackage{textcomp}
\usepackage{arydshln} 
\usepackage{amssymb}
\usepackage{pifont}

\newcommand{\xmark}{\ding{55}}%
\newcommand{\cmark}{\ding{51}}%

\usepackage{colortbl}
\usepackage{xcolor}
\definecolor{mygrey}{RGB}{240,240,240}
\definecolor{mydarkgrey}{RGB}{220,220,220}

\title{Leveraging User-Generated Metadata of Online Videos for Cover Song Identification}

\author{Simon Hachmeier \and Robert Jäschke \\
        Berlin School of Library and Information Science \\ Humboldt-Universität zu Berlin \\ \texttt{\{simon.hachmeier,robert.jaeschke\}@hu-berlin.de}}

\begin{document}
\maketitle
\begin{abstract}
YouTube is a rich source of cover songs. Since the platform itself is organized in terms of videos rather than songs, the retrieval of covers is not trivial. The field of cover song identification addresses this problem and provides approaches that usually rely on audio content. However, including the user-generated video metadata available on YouTube promises improved identification results.
In this paper, we propose a multi-modal approach for cover song identification on online video platforms. We combine the entity resolution models with audio-based approaches using a ranking model.
Our findings implicate that leveraging user-generated metadata can stabilize cover song identification performance on YouTube.  
\end{abstract}


\definecolor{ffi}{rgb}{0.998,0.722,0.635} 
\definecolor{sha}{rgb}{0.635,0.998,0.722} 

\definecolor{rja}{rgb}{0.878, 0.831, 0.482} 
\definecolor{msc}{rgb}{0.721, 0.576, 0.862} 
\definecolor{far}{rgb}{0.576, 0.862, 0.592} 
\definecolor{hsa}{rgb}{0.576, 0.788, 0.862} 
\definecolor{mpa}{rgb}{0.978,0.534,0.534} 

\definecolor{TODO}{rgb}{0.784,0.145,0.00}

\newcommand{\pkomm}[3][TODO]{\todo[color=#1,size=\scriptsize,#2]{\sffamily #1: #3}}

\newcommand{\ffi}[1]{\pkomm[ffi]{inline}{#1}}\newcommand{\mffi}[1]{\pkomm[ffi]{noinline}{#1}} 
\newcommand{\sha}[1]{\pkomm[sha]{inline}{#1}}\newcommand{\msha}[1]{\pkomm[sha]{noinline}{#1}} 
\newcommand{\rja}[1]{\pkomm[rja]{inline}{#1}}\newcommand{\mrja}[1]{\pkomm[rja]{noinline}{#1}} 
\newcommand{\msc}[1]{\pkomm[msc]{inline}{#1}}\newcommand{\mmsc}[1]{\pkomm[msc]{noinline}{#1}} 
\newcommand{\far}[1]{\pkomm[far]{inline}{#1}}\newcommand{\mfar}[1]{\pkomm[far]{noinline}{#1}} 
\newcommand{\hsa}[1]{\pkomm[hsa]{inline}{#1}}\newcommand{\mhsa}[1]{\pkomm[hsa]{noinline}{#1}} 
\newcommand{\mpa}[1]{\pkomm[mpa]{inline}{#1}}\newcommand{\mmpa}[1]{\pkomm[mpa]{noinline}{#1}} 

\newcommand{\final}[1]{\textbf{/* for camera ready/long version: #1  */}}

\newcommand{\bti}{\tau}
\newcommand{\barti}{\alpha}
\newcommand{\basetitle}{\textrm{base}_\bti}
\newcommand{\baseartisttitle}{\textrm{base}_\barti}

\newcommand{\indi}{I} 
\newcommand{\univ}{U} 

\newcommand{\eg}{e.g.,\xspace}
\newcommand{\ie}{i.e.,\xspace}

\newcommand{\sh}{SHS\xspace}
\newcommand{\shs}{SecondHandSongs\xspace}
\newcommand{\shsK}{SHS100K\xspace}
\newcommand{\vshs}{V-SHS\xspace}
\newcommand{\datacos}{DaTacos\xspace}
\newcommand{\dat}{DaT\xspace}
\newcommand{\vdat}{V-DaT\xspace}
\newcommand{\shsyt}{SHS-YT\xspace}
\newcommand{\shsseed}{SHS-SEED\xspace}
\newcommand{\ytcrawl}{YT-CRAWL\xspace}

\newcommand{\ytb}[1]{\href{https://youtu.be/#1}{#1}}

\setcounter{topnumber}{5}
\setcounter{bottomnumber}{5}
\setcounter{totalnumber}{20}
\renewcommand{\topfraction}{1}
\renewcommand{\bottomfraction}{1}
\renewcommand{\textfraction}{0}
\renewcommand{\floatpagefraction}{0.99}



\section{Introduction}
\label{sec:intro}

\sha{RV 1: The paper is dense with many acronyms, especially given the variety of datasets used, making it difficult to follow at times. Despite this, the experiments are sound, and the topic is interesting.}
\sha{RV 2: The writing is good. But it would be great to have a short and clear description / illustration of the overall system. It took me multiple times of reading back and forth to fully understand how the whole system works.}



Music is a popular content category on YouTube \cite{montero2020digital}. 
Uploaders share music in a variety of contexts, ranging from amateur covers to mashups \cite{follow_the_algorithm_music, music_youtube_user_engagement}. Since YouTube is not organized in terms of songs but rather in terms of online videos,\footnote{Except for YouTube's streaming service \emph{YouTube Music}.} finding cover versions of songs is a non-trivial retrieval task. 

Driven by applications such as copyright infringement detection, cover song identification (CSI) deals with the retrieval of covers. The key challenge of CSI is to compare songs based on properties which can indicate their association (\eg melody, lyrics) while discarding irrelevant information (\eg timbre). Consequently, current research efforts are mainly audio-based \cite{Du2023bytecover3, liu2023coverhunter, hu2022lyracnet, yu2020cqtnet} with limited consideration of non-audio features such as lyrics  \cite{what_if_two_musical_versions}.
However, the utilization of user-generated metadata of online videos like in similar tasks \cite{copyright_infringement_detection, classifying_derivative_works}, has not yet been considered in CSI.

We propose to model CSI on online video platforms (OCSI) as a multi-modal problem, based on the hypothesis that uploaders tend to describe their videos using attributes of songs (\eg song title, performer name) to make them easily findable. In this work, we propose multi-modal ensembles combining entity resolution models with audio-based CSI models in a late-fusion fashion using the ranking model LambdaMART \cite{wu2010Adaptingboostinginformationretrievalmeasures}. We compare the performance of the proposed ensembles with the performance of the CSI models. Further, we study the robustness of ER approaches in difficult cases such as song title variations or hard negatives on YouTube and provide our code and results.\footnote{\url{https://github.com/progsi/er_csi}}

\definecolor{colshs}{rgb}{.447,.624,.812}
\definecolor{colyt}{rgb}{.788,.129,.118}
 
\begin{figure*}
  \centering
  \includegraphics[scale=0.3]{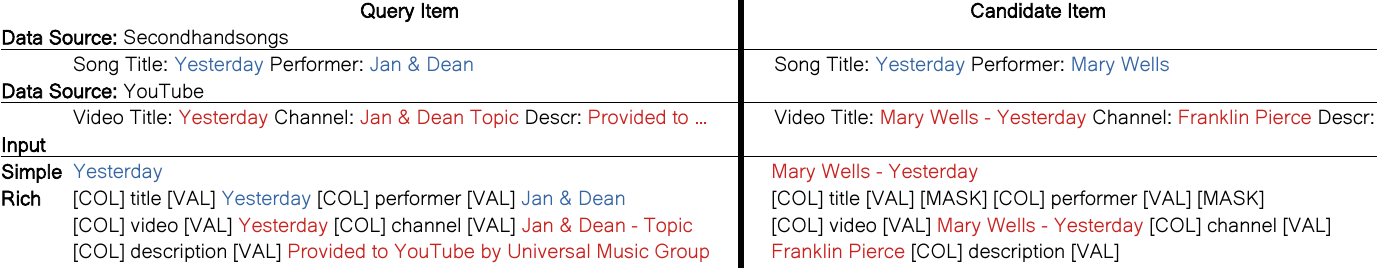}
  \caption{Example of the input of items of the work ``Yesterday'' written by John Lennon and Paul McCartney. 
  Colors in the box frames and text indicate the data source: \textcolor{colshs}{blue} stands for \textcolor{colshs}{Secondhandsongs} and \textcolor{colyt}{red} for \textcolor{colyt}{YouTube}.}
  \label{fig:text_input}
\end{figure*}

\section{Multi-Modal Online Cover Song Identification}\label{sec:methods}


The items in our task are online videos. 
The goal of OCSI is the retrieval of items associated with the same musical work as a query item. In traditional CSI, only representations concerning musical content (\eg audio and lyrics) are considered. We additionally leverage user-generated metadata. In Figure~\ref{fig:text_input} we show an example of a query-candidate pair. 

Each item is represented by attributes derived from its audio data and attributes from its user-generated metadata (video title, channel name, video description, and a set of keywords). For simplicity, we assume that each item contains only one song. A song has the attributes song title, performer name, and a work identifier. Songs which are associated with the same musical work are considered relevant in the retrieval scenario when one song is used as a query to retrieve the other song. 

We model the task of OCSI as a multi-modal problem involving metadata and audio representations. Our overall system will consist of modules to compute the pairwise similarities for each of both modalities. Then, a ranking model combines both outputs to compute an overall rank.  
In the following, we explain our entity-resolution (ER) methods used to model similarities in the metadata domain.


\subsection{Entity Resolution}

\paragraph{Fuzzy Matching} We use the token ratio function from rapidfuzz \cite{bachmann2021maxbachmann}, which turned out to be the best performing fuzzy matcher for the task in a preliminary experiment. For a pair of strings, the function returns the maximum between their normalized Indel similarity and the token set ratio.
The former is the minimum number of insertions and deletions to convert one string into the other. 
The latter is the number of tokens in the shorter string contained in the longer string divided by the number of tokens in the longer string.
We validated the use of the token ratio by the MAP on the validation dataset (cf. Section~\ref{sec:experiment}).
We tested matching song title and performer concatenated and using solely the song title of the query item. For the candidate item, we experimented with only the video title and with the latter combined with the other attributes concatenated by space. The best configuration was simply matching the song title to the video title (cf. \emph{Simple} input in Figure~\ref{fig:text_input}). We also experimented with the snowball stemmer from NLTK \cite{teamnltkNaturalLanguageToolkit} for all attributes but it did not improve the results. 


\paragraph{S-BERT}\label{sec:sbert}


The model S-BERT \cite{reimers2019sentence} addresses the problem of quadratic complexity of language models (LMs) like BERT \cite{bert_original} when processing pairs of sentences. Hence, it was used as promising method for ER \cite{li2021improving, paganelli2023multi}. The model learns to encode sentences into embedding vectors which can be compared using similarity measures such as Cosine similarity. We fine-tune a multilingual approach of S-BERT \cite{reimers2020making} which encodes text sequences into 384-dimensional vectors. Like for fuzzy matching, we use the \emph{Simple} input because it performed better than the other attribute combinations. 
We apply a similar training procedure like in recent CSI approaches: we use triplet loss with a margin of 0.3 and apply online hard triplet mining \cite{xuan2020hard} where the hardest triplets in the batch of 16~items with 4~random works represented by 4~items are used for training updates. We select the best model after 10 epochs measured in MAP on the validation dataset.

\paragraph{Ditto}\label{sec:er_lm}

Other state-of-the-art approaches rely on contextualized embeddings provided by pretrained LMs to predict the matching confidence for a given entity pair. In theory, this can improve the ER task since the context is not only considered for tokens in one entity but across both entities. Thus, we experimented with Ditto \cite{Li2020DeepEntity}, HierGAT \cite{Yao2022EntityResolution}, and r-SupCon \cite{Peeters2022SupervisedContrastive}. 
We found that Ditto was both -- better performing and faster in inference. We therefore select Ditto for our experiments. The model computes a binary matching confidence based on entity pairs encoded by LMs such as RoBERTa \cite{liu2019roberta}. Due to the quadratic complexity during inference, we use S-BERT as blocker. We adopt the top-$k$ blocking strategy suggested by the authors \cite{li2021deep} where the top-$k$ most similar candidate items per query item are passed to Ditto and the remaining pairs are predicted by the blocker.

As underlying LM, we chose RoBERTa since it was shown to achieve high performance in the ER task \cite{Li2020DeepEntity, Peeters2022SupervisedContrastive, peeters2023wdc, Yao2022EntityResolution} and multilingual variant of BERT (mBERT) \cite{bert_original}. 
First, we experimented with the same attribute combinations we use for fuzzy matching and S-BERT. We observed that Ditto works better when items are represented with the same attributes on the query and candidate side. Hence, we use a \emph{rich input} (cf. Figure~\ref{fig:text_input}). As inputs for the LMs, we concatenate the names and values of attributes of the query and candidate item: Each attribute is represented by a [COL] token followed by its name (\eg \emph{title} for the song title) and a [VAL] token followed by its value (\eg \emph{Yesterday}). Since the actual song attributes of the items on the candidate side are not known, we mask the respective tokens of those using a [MASK] token. 
We fine-tune with a batch size of 32 and a learning rate of 1e-05 and a sequence length of 256 tokens. We select the best performing model after 15~epochs. We fine-tune Ditto with a dataset of pairs (cf. Section~\ref{sec:experiment}) and measure the performance on the validation dataset in F1.




\subsection{Combining ER with CSI}

We form multi-modal ensembles each combining one fine-tuned ER model with a trained CSI model (ER-CSI ensembles). We use two pre-trained CSI models: CQTNet \cite{yu2020cqtnet} and CoverHunter \cite{liu2023coverhunter}. Both models encode items into vectors and represent musical similarity using Cosine similarity. The former uses convolutional neural networks to learn 300-dimensional vector representations. CoverHunter uses conformer neural networks \cite{gulati2020conformer} and an attention mechanism for temporal pooling \cite{attentive_statistics_pooling} to learn 128-dimensional vector representations. For each ER-CSI ensemble we train the ranking model LambdaMART \cite{wu2010Adaptingboostinginformationretrievalmeasures} using the pairwise similarities as input features. We use (mean average precision) MAP  objective function and consider the top 50 feature interactions similar to \cite{lucchese2022ilmart}.

\section{Experimental Setup}\label{sec:experiment}

Our experiments aim to evaluate a) whether ensembles of ER and CSI models outperform CSI models and b) whether ER models are robust against hard negatives and song title variations (\eg translations of song titles or parodies).\footnote{An example for a parody title is ``Bye, Bye Johnny'' by The Rattles covering ``Johnny B. Goode''} We report two evaluation metrics suggested by MIREX:\footnote{cf. \url{https://www.music-ir.org/mirex/wiki/2021:Audio_Cover_Song_Identification}} MAP and mean rank of the first relevant item (MR1). 

We use subsets of two popular CSI datasets: \shsK \cite{key_invariant_shs100k2} and \datacos \cite{datacos_yesiler2019}, which are popular in CSI research. 
Both datasets contain items which are songs represented by metadata attributes, a YouTube identifier, and a work identifier. We only retain items with available videos and denote the resulting datasets as \vshs and \vdat respectively (cf. Table~\ref{tab:datasets}).
We retrieve YouTube metadata for all the videos using \emph{YouTube Search Python} and extract the audio features as described by the authors of CoverHunter \cite{liu2023coverhunter} and CQTNet \cite{yu2020cqtnet}.



\paragraph{Training and Validation Subsets}\label{sec:train_val_set}


We fine-tune S-BERT and Ditto and train LambdaMART using a random sample of items from the V-SHS100K training set with 1,000 positive pairs and 6,000 negative pairs similar to datasets by \citet{konda2016magellan}.
Similarly, we create a pair-wise validation dataset used to select the best model checkpoint of Ditto. The full overview of datasets used is given in Table~\ref{tab:datasets}.


\begin{table}
  \centering
  \begin{tabular}{>{\kern-\tabcolsep}llrrccc<{\kern-\tabcolsep}}
    \hline
    Dataset & Split & Works & Avg. & FM & DI & SB \\
    \hline
    \vshs-F & Train &  1,501 & 1.33 & \xmark & \cmark
 & \cmark
\\
    \vshs-V & Valid & 1,827 & 4.73 & \cmark & \xmark  & \cmark \\
    \vshs-P & Valid & 1,224 & 1,63  & \xmark & \cmark
& \xmark \\
    \rowcolor{mydarkgrey}
    \vshs-T & Test & 1,679 & 5.25 & \cmark
 & (\cmark) & \cmark
\\
     \rowcolor{mydarkgrey}
     \rowcolor{mydarkgrey}
     -Unique & Test & 852 & 2.75 & \cmark
 & \cmark
 & \cmark
 \\
 \rowcolor{mydarkgrey}
     -Noise & Test & 12 & 4.15 & \cmark
 & \cmark
 & \cmark
 \\
     \rowcolor{mygrey}
    V-\dat & Test & 2,784 & 4.92 & \cmark
 & (\cmark)  & \cmark \\
  \hline
  \end{tabular}
  \caption{Dataset statistics: the number of works (Works) with average number of items (Avg.) and usage for Fuzzy Matching (FM), Ditto (DI), or S-BERT (SB). (\cmark) denotes partial use (after blocking).}
  \label{tab:datasets}
\end{table}

\paragraph{Test Subsets}

Addressing a), we select the \emph{V-\dat} and \emph{\vshs-T} which are subsets with available videos of datasets typically used in CSI evaluation. For b), the robustness study, we create additional test datasets. First, we only retain one item per work and song title (dropping multiple items with the same song title per work). This subset based on \emph{\vshs-T} is denoted by \emph{\vshs-T-Unique}.

Lastly, we aim to evaluate the ER models' robustness to hard negatives. On YouTube, this can be expected in cases where either the song title is the same for different works or when the words in the song title are used in a different context (\eg the song title ``Hush'' occurring in the sentence ``Relaxing Hush Sounds''). We focus on the latter problem which is particularly challenging for song titles with one word, because these words are more likely to occur in different contexts. We create a subset of \emph{\vshs-T} containing only items with one-word song titles. For each of the works in the dataset we instruct ChatGPT~3.5 \cite{ChatGPT} to generate video metadata containing the respective word.\footnote{The prompt was: \emph{Consider the following a list of words. Generate a meaningful video title for each of these words. }} We show some examples in Table~\ref{tab:one_word_song_titles}. The resulting dataset \emph{\vshs-T-Noise} contains 12~works each with 5 generated video titles (hard negatives) and on average around 4~items of the work.

\begin{table}
    \centering
\small    
\begin{tabular}{@{}lp{50mm}@{}}
\hline
Song Title &  Utterances in generated video title \\
\hline
Yesterday & \textbf{Yesterday}'s Kitchen: Old Recipes
\\
Hush & Relaxing \textbf{Hush} Sounds
\\
Time & Mastering \textbf{Time} Management
 \\
\hline
\end{tabular}
    \caption{Examples of one-word song titles and video titles of generated  hard negatives with ChatGPT~3.5 in the SHS100K-T-Noise dataset.} \label{tab:one_word_song_titles}
\end{table}





\section{Results}\label{sec:results}





In Table~\ref{tab:overall_results} we report the results of  experiment a). Generally, we observe a strong improvement in MR1 when comparing ER-CSI ensembles with CSI. Improvements in MAP are evident but smaller for ensembles with CoverHunter. Still, gains up to +9\% in MAP and up to -13.45 ranks in MR1 are achieved on V-DaTacos. The ensembles with Fuzzy Matching have relatively small performance gains and do not achieve a higher MAP than CoverHunter.

The highest results in MAP are achieved with S-BERT and Ditto with S-BERT as blocker and RoBERTa as underlying LM. Considering the computational overhead for using Ditto, even with blocking and $k=100$, makes its use for the task questionable. However, Table~\ref{tab:robustness_results} shows that combining Ditto with S-BERT can stabilize robustness in some cases. Even though the latter achieves higher MAP on the -Unique subsets, combining it with Ditto (M) yields +5\% and +8\% in MAP on those. Apparently, this might be due to better rankings of Ditto at the earlier ranks indicated by MR1. 

\begin{table}[h]
  \centering
  \begin{tabular}{>{\kern-\tabcolsep}lrrrrr<{\kern-\tabcolsep}}
    \hline
     & 
    \multicolumn{2}{c}{\textbf{\vshs-T}} & 
    \multicolumn{2}{c}{\textbf{\vdat}} \\
     & MAP & MR1 & MAP & MR1   \\
    \hline
    \rowcolor{mygrey}
    CQTNet & 0.71 & 47.40 &  0.74 & 10.74  \\
    -Fuzzy & 0.75 & 14.92 & 0.80 & 4.16  \\
    -S-BERT & \textbf{0.85} & \textbf{12.14} &  0.91 & 3.06  \\
    -SB+Ditto (R) & \textbf{0.85}  & 16.29 & \textbf{0.92} & \textbf{3.03} \\
    -SB+Ditto (M) & 0.83 & 20.62 & 0.90 & 3.49   \\
    \hline
    \rowcolor{mygrey}
    CoverHunter & 0.92 & 12.60 & 0.84  & 15.71 \\
    -Fuzzy & 0.90 & 4.47 & 0.84 & 5.57 \\
    -S-BERT & \textbf{0.93} & \textbf{3.58} & 
    \textbf{0.93} & 3.00  \\
    -SB+Ditto (R) & \textbf{0.93} & 5.41 & \textbf{0.93} & \textbf{2.26} \\
    -SB+Ditto (M) & 0.92 & 7.06 & 0.91 & 2.70 \\
    \hline
  \end{tabular}
   \caption{Experiment a): Performances of ER-CSI ensembles against CSI models. SB+Ditto denotes Ditto with S-BERT as blocker; (R) stands for RoBERTA and (M) for mBERT. }
     \label{tab:overall_results}
\end{table}

\begin{table}[h]
  \centering
  \begin{tabular}{@{}lrrrr@{}}
    \hline
      & \multicolumn{2}{c}{\textbf{-Noise}} & \multicolumn{2}{c}{\textbf{\sh-Uniq.}} \\ & MAP & MR1 & MAP & MR1 \\
    \hline
    Fuzzy & 0.38 & 5.03 & 0.37 & 189.78  \\
    S-BERT & 0.53 & 4.15 & 0.55 & 138.04 \\
    Ditto (R) & 0.43 & \textbf{2.97} & 0.37 & 114.94 \\
    Ditto (M) & 0.49 & 4.21 & 0.46 & \textbf{100.26} \\
    SB+Ditto (R) & \textbf{0.56} & \textbf{2.97} & 0.51 & 183.22  \\
    SB+Ditto (M) & 0.44 & 7.31 & \textbf{0.60} & 252.31  \\
    \hline
  \end{tabular}
    \caption{Experiment b): Results of ER approaches on the \vshs-T-Noise (-Noise), \vshs-T-Unique (S-Uniq.). Due to the smaller dataset size, we set $k=10$ for -Noise.} \label{tab:robustness_results}

\end{table}

\section{Limitations and Conclusion}
\label{sec:discussion}

In this paper, we implemented ER approaches as means to support the task of cover song identification in online videos on the example of the online video platform YouTube. We showed that simple fuzzy matching can partly help to increase model performances. Better results were achieved by using S-BERT. Additionally, using Ditto appears to be adequate only in some cases such as song title variations.
However, the robustness of ER models is generally harmed by hard negatives which potentially do not refer to music. 

Lastly, we outline limitations of this study. Our selected text input structure for S-BERT and fuzzy matching can only detect the song title in the video titles. While this might be sufficient for our utilized datasets, other videos which contain the title only in the description or keywords are not uncovered. Secondly, song titles that are completely different than the reference (\eg parodies or medleys) cannot be detected by ER models. Hence, we see these models as supporting tool in OCSI rather than independent approaches.
In future research, we aim to 
leverage more recent large LMs which are starting to get used for ER \cite{peeters2024entity}. 



\bibliography{literature}




\end{document}